\documentclass[aps,prx,groupedaddress,twocolumn,longbibliography]{revtex4-1}%\documentclass[aps,pra,twocolumn,amsmath,superscriptaddress,floatfix,longbibliography]{revtex4-1}
\usepackage{amsmath}
\usepackage{epsfig}
\usepackage{graphicx}
\usepackage{graphics}
\usepackage{url}
\usepackage[english]{babel}
\usepackage{dcolumn}
\usepackage{braket}
\usepackage{color}
\usepackage{hyperref}

\newcommand{\cre}[1]{\hat{#1}^{\dagger}}

\begin{document}
\title{Theory of absorption lineshape in monolayers of transition metal dichalcogenides}

\author{F.~Lengers}
\affiliation{Institut f\"ur Festk\"orpertheorie, Universit\"at M\"unster,
Wilhelm-Klemm-Str.~10, 48149 M\"unster, Germany}

\author{T.~Kuhn}
\affiliation{Institut f\"ur Festk\"orpertheorie, Universit\"at M\"unster,
Wilhelm-Klemm-Str.~10, 48149 M\"unster, Germany}

\author{D.~E.~Reiter }
\affiliation{Institut f\"ur Festk\"orpertheorie, Universit\"at M\"unster,
Wilhelm-Klemm-Str.~10, 48149 M\"unster, Germany}

\date{\today}

\begin{abstract}
The linear absorption spectra in monolayers of transition metal dichalcogenides show pronounced signatures of the exceptionally strong exciton-phonon interaction in these materials. To account for both exciton and phonon physics in such optical signals, we compare different theoretical methods to calculate the absorption spectra using the example of $\mathrm{MoSe_2}$.  
In this paper, we derive the equations of motion for the polarization either using a correlation expansion up to 4th Born approximation or a time convolutionless master equation. We show that the Born approximation might become problematic when not treated in high enough order, especially at high temperatures. In contrast, the time convolutionless formulation gives surprisingly good results despite its simplicity when compared to higher-order corrrelation expansion and therefore provides a powerful tool to calculate the lineshape of linear absorption spectra in the very popular monolayer materials.
\end{abstract}

%\pacs{78.67.Hc; 78.47.D-; 42.50.Md}

\maketitle

\section{Introduction}
Monolayers of transition metal dichalcogenides (TMDCs) are intensely studied direct semiconductors with tightly bound excitons \cite{Mak10,Chernikov14,Steinleitner17,Wang18,Mueller18} and exceptionally strong exciton-phonon interaction \cite{Christiansen17,Chow17,Shree18,Glazov19}. Remarkable signatures of these traits are strongly inhomogeneous lineshapes in absorption spectra and pronounced phonon-assisted transitions in photoluminescence \cite{Christiansen17,Chow17,Shree18,Niehues18}. This gives rise to the question how to theoretically describe the combined exciton and phonon physics. In this paper, we will discuss the correlation expansion approach, which is well established in III-V semiconductors like GaAs, and furthermore promote a time convolutionless (TCL) master equation approach to calculate linear optical signals in TMDC monolayers.

In III-V semiconductor structures like GaAs the electron-phonon interaction is rather weak. These structures have been investigated with a range of well established theoretical methods. One common method is the correlation expansion approach, which treats the infinite hierarchy of equations of motion using the correlations of higher-order density matrices \cite{Rossi02}, which also has been applied to III-V nanostructures like quantum dots \cite{Krummheuer02, Foerstner03,Kruegel06}, quantum wires \cite{Reiter07} or quantum well structures \cite{Binder98,Weber09}. This method relies on the idea that higher-order correlations can be neglected due to the weak electron-phonon interaction.

Therefore, it is unclear, if this approach will hold for the case of TMDCs and which order in the expansion is required. 
Existing theoretical descriptions of optical signals primarily focus on the fundamental excitonic resonances by solving a simplified Wannier model for the correlated electron-hole pairs \cite{Berghaeuser14,Malic18,Meckbach18} or the Bethe-Salpeter equation within ab-initio methods \cite{Qiu16,Deilmann17,Drueppel18}. There the lineshape of the excitonic resonance is usually incorporated phenomenologically as a Lorentzian. Because the origin of lineshapes are diverse \cite{Moody15}, a few works are devoted to the description of the influence of exciton-phonon interaction on spectra \cite{Selig16,Christiansen17,Shree18}. In Refs.~\cite{Selig16,Christiansen17} a quantum kinetic model is used where the line broadening is self-consistently calculated within a Born-Markov approximation. The non-Markovian inhomogeneous lineshape due to phonons is then calculated by exploiting the 2nd Born approximation where all quanities are damped by the self-consistently calculated broadenings. This results in a good agreement with experiment and hints at the fact that the 2nd Born approximation alone is not sufficient to calculate absorption spectra such that a detailed analysis of higher-order correlation expansion is of interest. In Ref.~\cite{Shree18} the exciton-phonon interaction is treated by retarded Green functions with a self-consistently calculated exciton self energy in the case of interaction with acoustic phonons which results in a similar form of the absorption spectrum as in Ref.~\cite{Christiansen17}.  
Here, we will critically discuss different levels of correlation expansion and show that in the case of TMDCs low orders are insufficient to correctly describe the spectra. To be specific, we will show that due to the strong exciton-phonon interaction the truncation on the 2nd or 4th Born approximation results in artifacts in the spectra. In contrast, the TCL master equation approach will be shown to be free of these problems and might therefore be better suited for the description of linear optical signals in TMDCs. 

To quantify our methods we will calculate the optical absorption spectrum of a TMDC monolayer using different levels of approximation
	\begin{enumerate}
	\itemsep0em
	\item Correlation expansion within a 2nd/4th Born approximation (2BA/4BA)
	\item Correlation expansion within a damped 4th Born approximation (4BA-D)
	\item Time Convolutionless Master Equation (TCL)
	\end{enumerate}
The comparison of the results also with existing experimental measurements \cite{Christiansen17,Shree18} shows that the TCL approach has many advantages compared to the correlation expansion for the calculation of linear spectra. The derived damped version of correlation expansion (4BA-D) will also show why the self-consistently damped Born approximation in Ref.~\cite{Christiansen17} compares so well with experiment.
	
The remainder of the paper is organized as follows: We start by introducing the Hamiltonian and the different levels of approximation in Sec.~\ref{sec:theory}. We then present our results in Sec.~\ref{sec:results}. In particular we perform a numerical study of the absorption spectra using the different methods in Sec.~\ref{sec:results:abs}. We furthermore analytically compare the equations for the different approaches in Sec.~\ref{sec:results:comparison}. 
We then briefly discuss the case of weak coupling in Sec.~\ref{sec:results:weak}, before concluding in Sec.~\ref{sec:concl}.
\section{Theory} \label{sec:theory}
\begin{figure}[ht]
\includegraphics[width=\columnwidth]{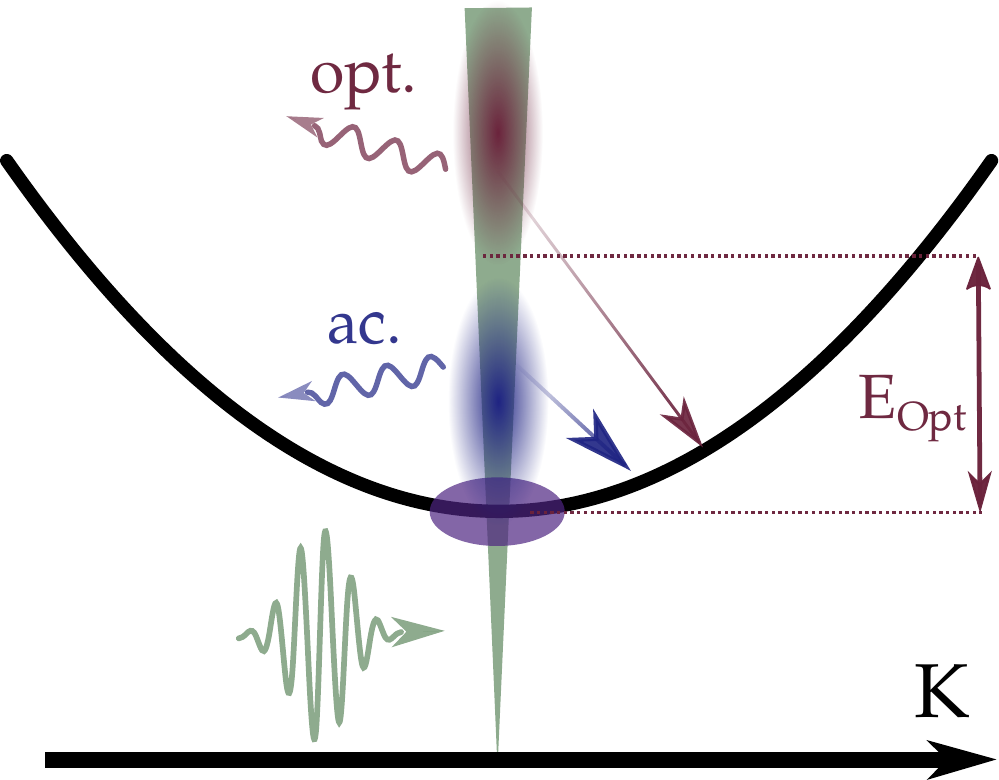}
\caption{Sketch of our system: We consider the $1s$-exciton parabola, where the $\textbf{K}=0$ exciton (violet dot) lies within the light cone (green area). Phonon-assisted processes by optical phonons (red) and acoustic phonons (blue) lead to transitions above the optical bandgap for $T=0$~K. The shaded areas indicate where the respective phonons are efficient.
}
\label{fig_introduction}
\end{figure}
Starting point of our considerations is the Hamiltonian in the exciton picture for TMDCs \cite{Selig16,Christiansen17,Katsch18} as also sketched in Fig.~\ref{fig_introduction}
\begin{align}
	H=&\sum\limits_{\mathbf{K}}E_{\mathbf{K}}\cre{P}_{\mathbf{K}}\hat{P}_{\mathbf{K}}+\sum\limits_{\mathbf{K},i}\hbar\omega_{\mathbf{K},i}\cre{b}_{\mathbf{K},i}\hat{b}_{\mathbf{K},i} \nonumber \\
	&-M_{1s}E(t)\cre{P}_0-\left(M_{1s}E(t)\right)^*\hat{P}_0 \nonumber \\
	&+\sum\limits_{\mathbf{K},\mathbf{Q},i}g_{\mathbf{Q},i}\cre{P}_{\mathbf{K}+\mathbf{Q}}\hat{P}_{\mathbf{K}}\left(\hat{b}_{\mathbf{Q},i}+\cre{b}_{-\mathbf{Q},i}\right) \,.
	\label{eq:Hamiltonian}
\end{align} 

The first term describes the $1s$ exciton parabola with $\cre{P}_{\mathbf{K}}$ and  $\hat{P}_{\mathbf{K}}$ being the exciton creation and annihilation operator at the center-of-mass momentum $\mathbf{K}$ at the $K$-valley with energy dispersion $E_{\mathbf{K}}=E_{1s}+\frac{\hbar^2}{2M}K^2$ with the exciton mass $M$ and bound state energy $E_{1s}$. Details on the exciton operator and material parameters can be found in Appendix~\ref{app_mat_parameters}. For the sake of clarity we focus on the linear absorption of Mo-based TMDCs such that intervalley-processes are of minor importance because in these materials the other optically dark excitonic valleys lie energetically above the bright one \cite{Selig16,Malic18}. We therefore restrict ourselves to one valley for the linear absorption. 
The second term is the free phonon Hamiltonian given by the phonon creation and annihilation operator $\cre{b}_{\mathbf{K},i}$ and $\hat{b}_{\mathbf{K},i}$  of branch $i$, momentum $\mathbf{K}$ and energy $\hbar\omega_{\mathbf{K},i}$.
The third term describes the light-matter interaction in the usual dipole approximation, where we consider right-circularly polarized light such that the exclusive excitation of the $K$ valley is a good approximation \cite{Wang18}. The dipole matrix element  projected onto the polarization of the light field $\mathbf{e}_0$ is denoted by $M_{1s}=\mathbf{M}_{1s}\cdot\mathbf{e}_0$ and $E(t)$ is the electric field amplitude which is only coupled to the optically active $\mathbf{K}=0$ exciton.
The last term describes then the exciton-phonon interaction with the intravalley exciton-phonon matrix element $g_{\mathbf{Q},i}$ \cite{Siantidis01,Selig16}. We consider both acoustic and optical phonons which mainly couple in deformation potential approximation with one optical branch at $E_{\mathrm{opt}}=34.4\,$meV and one effective acoustic branch with the sound velocity $c_s=4.1~\mathrm{nm/ps}$. All material parameters alongside a brief derivation of the exciton-phonon interaction are given in Appendix~\ref{app_mat_parameters}. 

Here, we are interested in the absorption spectrum $\alpha(\omega)$, which in the case of excitation by an ultrafast optical pulse is proportional to the Fourier transform of the polarization $\mathbf{P}(t)=\left( \mathbf{M}_{1s}^*\langle\hat{P}_0\rangle+\mathbf{M}_{1s}\langle \cre{P}_0\rangle\right)$. Assuming an excitation parallel to the dipole matrix element $\mathbf{M}_{1s}$ by a delta pulse in rotating wave approximation the absorption spectrum is calculated by
	\begin{align}
	\alpha(\omega)\propto \mathrm{Im}\left[\langle\hat{P}_0\rangle (\omega)\right]=:\mathrm{Im}\left[P_0(\omega)\right] \,.
	\end{align}
Considering the excitation by a delta pulse we set $P_0(t=0)=i$ as an initial condition considering that the pulse is so short that the exciton-phonon interaction only acts after the pulse. All correlations are initially set to $0$ and the spectrum is calculated by Fourier transform of the subsequent dynamics. 

For the phonons, we assume a thermalized phonon bath with a Bose distribution $n_{\mathbf{K},i}$ of branch $i$. We note that deviations from $n_{\mathbf{K},i}$ would only contribute to the dynamics in higher order of the electric field and are therefore neglected.\\

With this background, we can compute the equations of motion using the standard Heisenberg equations. Considering only the linear response to an electric field and an undoped system, we can neglect exciton-exciton correlations usually important for nonlinear optics \cite{Axt982,Katsch19} due to a low charge density. However, the many-body nature of the exciton-phonon interaction results in an infinite hierarchy of equations of motion. In the following we present different methods how to deal with this hierarchy.

\subsection{Quantum Kinetic equations in Born approximation}
\label{sec:QK_BA}
A general technique for obtaining a closed set of equations of motion is the correlation expansion \cite{Rossi02}. These equations are non-Markovian and truncate the correlations at a certain order usually called the corresponding Born approximation. Here, we are interested in the coherent exciton polarization $P_0=\langle\hat{P}_0\rangle$, for which the equation of motion up to linear order in the electric field reads
	\begin{align}
	i\hbar\frac{d}{dt}\langle\hat{P}_0\rangle=&
	\left(E_{1s}-\frac{i\hbar}{\tau}\right)\langle\hat{P}_0\rangle-M_{1s}E(t) \nonumber \\
	&+\sum\limits_{\mathbf{K},i}g_{\mathbf{K},i}\left(\langle\hat{P}_{-\mathbf{K}}\hat{b}_{\mathbf{K},i}\rangle +\langle\hat{P}_{-\mathbf{K}}\cre{b}_{-\mathbf{K},i}\rangle\right) \nonumber \\
	=:&\left(E_{1s}-\frac{i\hbar}{\tau}\right)P_0-M_{1s}E(t) \nonumber \\
	&+\sum\limits_{\mathbf{K}}\left(S_{\mathbf{K},i}^{(-)}+S_{\mathbf{K},i}^{(+)}\right). \label{eq:QK_P}
	\end{align}
The first line corresponds to the usual energy oscillation, where we introduced the dephasing rate of the $\mathbf{K}=0$ exciton by spontaneous emission $\tau$. The second line accounts for the influence of the phonons, where we abbreviate the correlations by $S_{\mathbf{K},i}^{(-)}=g_{\mathbf{K},i}\langle\hat{P}_{-\mathbf{K}}\hat{b}_{\mathbf{K},i}\rangle$ and $S_{\mathbf{K},i}^{(+)}=g_{\mathbf{K},i}\langle\hat{P}_{-\mathbf{K}}\cre{b}_{-\mathbf{K},i}\rangle$. These expectation values coincide with the correlations up to linear order in the electric field since $\langle \hat{b}_{\mathbf{K},i}\rangle$ is of 2nd order. However we would like to point out that the deformation potential coupling used here for optical phonons can lead to a non-vanishing coherent phonon amplitude for $\mathbf{K}=0$ unlike polar optical coupling \cite{Rossi02}. The correlations then obey the equations of motion
\begin{align}
i\hbar\frac{d}{dt}S_{\mathbf{K},i}^{(-)}=
&\left(E_{\mathbf{-K}}+\hbar\omega_{\mathbf{K},i}\right)S_{\mathbf{K},i}^{(-)}
\label{eom_Sminus}\\
&+\left|g_{\mathbf{K},i}\right|^2P_0\left(1+n_{\mathbf{K},i}\right)\nonumber
\\
&+\sum\limits_{\mathbf{K}',j}
\left(F_{\mathbf{K},\mathbf{K}'}^{(i,j)}+F_{\mathbf{K},\mathbf{K}'}^{(i,j),-}\right)\nonumber,
\end{align}
\begin{align}
i\hbar\frac{d}{dt}S_{\mathbf{K},i}^{(+)}=
&\left(E_{\mathbf{-K}}-\hbar\omega_{-\mathbf{K},i}\right)S_{\mathbf{K},i}^{(+)}
\label{eom_Splus}\\
&+\left|g_{-\mathbf{K},i}\right|^2P_0n_{-\mathbf{K},i}\nonumber\\
&+\sum\limits_{\mathbf{K}',j}
\left(F_{\mathbf{K},\mathbf{K}'}^{(i,j)+}+F_{\mathbf{K}',\mathbf{K}}^{(j,i)}\right)\nonumber,
\end{align}
where we defined
	\begin{align*}
	& F_{\mathbf{K},\mathbf{K'}}^{(i,j)}:=g_{\mathbf{K},i}g_{\mathbf{K'},j}\delta\langle \hat{P}_{-\mathbf{K-K'}}\cre{b}_{-\mathbf{K}',j}\hat{b}_{\mathbf{K},i}\rangle \, ,\\
	&F_{\mathbf{K},\mathbf{K'}}^{(i,j)+}:=g_{\mathbf{K},i}g_{\mathbf{K'},j}\delta\langle \hat{P}_{-\mathbf{K-K'}}\cre{b}_{-\mathbf{K}',j}\cre{b}_{-\mathbf{K},i}\rangle \, ,\\
	&F_{\mathbf{K},\mathbf{K'}}^{(i,j)-}:=g_{\mathbf{K},i}g_{\mathbf{K'},j}\delta\langle \hat{P}_{-\mathbf{K-K'}}\hat{b}_{\mathbf{K}',j}\hat{b}_{\mathbf{K},i}\rangle \, .
	\end{align*}
We denote $\delta\langle ...\rangle$ as the correlation defined as the expectation value subtracted by all lower-order factorizations \cite{Rossi02}, e.g., 
	\begin{align*}F_{\mathbf{K},\mathbf{K}'}^{(i,j)}=g_{\mathbf{K},i}g_{\mathbf{K}',j}
	&\left[\langle \hat{P}_{-\mathbf{K-K'}}\cre{b}_{-\mathbf{K}',j}\hat{b}_{\mathbf{K},i}\rangle\right.\\
	&\left.-\langle\hat{P}_{\mathbf{-K-K'}}\rangle\langle\cre{b}_{\mathbf{-K'},j}\hat{b}_{\mathbf{K},i}\rangle\delta_{i,j}\delta_{-\mathbf{K}',\mathbf{K}}\right],
	\end{align*} 
where we only considered factorizations which are linear in the electric field.

The \textit{2nd Born approximation}  (2BA) is obtained by neglecting $F,F^+,F^-$ in Eq.(\ref{eom_Sminus})-(\ref{eom_Splus}).

The \textit{4th Born approximation} (4BA) goes one step further and takes into account the equations of motion for $F,F^+,F^-$
\begin{widetext}
\begin{align}
	&i\hbar\frac{d}{dt} F_{\mathbf{K},\mathbf{K}'}^{(i,j)}=
	\left(E_{\mathbf{-K-K'}}+\hbar\omega_{\mathbf{K},i}-\hbar\omega_{-\mathbf{K}',j}\right) F_{\mathbf{K},\mathbf{K}'}^{(i,j)}	
	+|g_{\mathbf{K},i}|^2S_{\mathbf{K'},j}^{(+)}\left(1+n_{\mathbf{K},i}\right)
	+|g_{-\mathbf{K}',j}|^2S_{\mathbf{K},i}^{(-)} n_{\mathbf{-K}',j},
	\label{eq:QK_Fpm}
	\end{align}
	\begin{align}
	&i\hbar\frac{d}{dt}F_{\mathbf{K},\mathbf{K'}}^{(i,j)+}=
	\left(E_{\mathbf{-K-K'}}-\hbar\omega_{-\mathbf{K},i}-\hbar\omega_{-\mathbf{K}',j}\right)F_{\mathbf{K},\mathbf{K'}}^{(i,j)+}
	+|g_{\mathbf{-K},i}|^2S_{\mathbf{K}',j}^{(+)} n_{\mathbf{-K},i}
	+|g_{-\mathbf{K}',j}|^2S_{\mathbf{K},i}^{(+)} n_{\mathbf{-K}',j},
	\end{align}
	\begin{align}
	&i\hbar\frac{d}{dt}F_{\mathbf{K},\mathbf{K'}}^{(i,j)-}=
	\left(E_{\mathbf{-K-K'}}+\hbar\omega_{\mathbf{K},i}+\hbar\omega_{\mathbf{K}',j}\right)F_{\mathbf{K},\mathbf{K'}}^{(i,j)-}
	+|g_{\mathbf{K},i}|^2S_{\mathbf{K}',j}^{(-)}\left(1+n_{\mathbf{K},i}\right)
	+|g_{\mathbf{K}',j}|^2S_{\mathbf{K},i}^{(-)} (1+n_{\mathbf{K}',j}),
	\end{align}
\end{widetext}
where all 4-operator expectation values have been factorized into all lower-order factorizations and 4-operator correlations are neglected. 

We emphasize that in the n-th order Born approximation, the full equations of motion are exact up to n-th order in the exciton-phonon coupling.
\\
\subsection{Damped Born approxmation}
In order to account for higher-order correlations going beyond the n-th order correlation expansion, it is possible to include these using a Born-Markov approximation \cite{Schilp94}. We show this at the example of $F$. For this the equation of motion in Eq.(\ref{eq:QK_Fpm})  is formally integrated. We use the zeroth order Born approximation
	$$S_{\mathbf{K},i}^{(\pm)}(t-\tau)\approx e^{-\frac{i}{\hbar}\left(E_{-\mathbf{K}}\mp \hbar\omega_{\mp\mathbf{K},i}\right)(-\tau)}{S}_{\mathbf{K},i}^{(\pm)}(t)$$
 in the spirit of a slowly-varying dynamics in the interaction picture. The expression for $F$ then reads
\begin{align}
	F_{\mathbf{K},\mathbf{K}'}^{(i,j)}\approx
	{\gamma}_{\mathbf{K},i,\mathbf{K}'}^{(-)}(t)S_{\mathbf{K}',j}^{(+)}(t)
	+{\gamma}_{\mathbf{K'},j,\mathbf{K}}^{(+)}(t)S_{\mathbf{K},i}^{(-)}(t),
	\label{eq_4BA_approx}
	\end{align}
with 
	\begin{align}
	{\gamma}_{\mathbf{K},i,\mathbf{K}'}^{(\mp)}(t)=
	&-\frac{i}{\hbar}\left|g_{\pm\mathbf{K},i}\right|^2\left(\frac{1}{2}\pm\frac{1}{2}+n_{\pm\mathbf{K},i}\right)
	\label{eq_def_gamma}\\
	&\cdot\int_0^t e^{-\frac{i}{\hbar}\left(E_{-\mathbf{K}-\mathbf{K}'}-E_{-\mathbf{K}'}\pm \hbar\omega_{\pm\mathbf{K},i}\right)\tau}d\tau. \nonumber
	\end{align}
We here choose $t_0=0$ as the time of optical generation. From Eq.~\eqref{eq_4BA_approx} we clearly see that the 3-operator correlations couple all 2-particle correlations. Reinserting this back into the equation of motion for $S_{\mathbf{K},i}^{(\pm)}$ we obtain terms in the form of (here for example for Eq.~\eqref{eom_Sminus}) 
	\begin{align*}
	i\hbar\frac{d}{dt}S_{\mathbf{K},i}^{(-)}=...+\sum\limits_{\mathbf{K}',j}\left({\gamma}^{(+)}_{\mathbf{K}',j,\mathbf{K}}S_{\mathbf{K},i}^{(-)}+{\gamma}_{\mathbf{K},i,\mathbf{K}'}^{(-)}S_{\mathbf{K}',j}^{(+)}\right).
	\end{align*}
The term $\sim\gamma^{(+)}$ describes a time-dependent relaxation and renormalization for $S_{\mathbf{K},i}^{(-)}$, because the summation over $\mathbf{K}'$ does not contain any dynamical variables. Using a Random Phase Approximation (RPA) the term $\sim\gamma^{(-)}$ can be neglected, because $S_{\mathbf{K}',j}^{(+)}$ is an oscillating function of $\mathbf{K}'$. Next, we also perform the integration and approximation of $F^{-}$, such that the full equation of motion for  $S_{\mathbf{K},i}^{(-)}$ reads
\begin{align}
i\hbar\frac{d}{dt}S_{\mathbf{K},i}^{(-)}=
&\left(E_{\mathbf{-K}}+\hbar\omega_{\mathbf{K},i}\right)S_{\mathbf{K},i}^{(-)}
\label{eom_Sminus_D}\\
&+\left|g_{\mathbf{K},i}\right|^2P_0\left(1+n_{\mathbf{K},i}\right)\nonumber
\\
&-i\hbar{\Gamma}_{\mathbf{K}}S_{\mathbf{K},i}^{(-)}. \nonumber
\end{align}
with 
	\begin{align}
	\Gamma_{\mathbf{K}}(t)=\frac{i}{\hbar}\sum\limits_{\mathbf{K}',j}\left({\gamma}_{\mathbf{K'},j,\mathbf{K}}^{(-)}(t)+{\gamma}_{\mathbf{K}',j,\mathbf{K}}^{(+)}(t)\right)
	\label{eq:Gamma}
	\end{align}
and analogous for $S^{(+)}_{\mathbf{K}}$. Note that $\Gamma_{\mathbf{K}}$ is a complex function, therefore it contains both energy renormalizations \cite{Rossi02} as well as a damping \cite{Schilp94}. We call this level of approximation
\textit{damped 2nd order Born approximation} (2BA-D).

It is illustrative to perform the full Markov approximation to see that $\Gamma_{\mathbf{K}}$ indeed results in a damping rate. Following Ref.~\cite{Schilp94}, we integrate over the exponential function in Eq.~\eqref{eq_def_gamma} using the Markov approximation and extending the integration to $t\rightarrow\infty$
\begin{align*}\lim\limits_{t\rightarrow\infty}\left(-\frac{i}{\hbar}\right)\gamma_{\mathbf{K},i,\mathbf{K}'}^{(\pm)}(t)\propto \delta\left(E_{\mathbf{K'}}-E_{-\mathbf{K-K'}}\pm\hbar\omega_{\mp\mathbf{K},i}\right),
\end{align*}
where we omitted the imaginary part describing an energy renormalization.
In the equation of motion, due to $\Gamma_{\mathbf{K}}(t) S_{\mathbf{K},i}^{(-)} \to \gamma_0 S_{\mathbf{K},i}^{(-)}$ this results in a constant relaxation rate $\gamma_0$. Therefore, we can interpret this as a damping of the correlations due to higher-order correlations. Note that for the results in Sec.~\ref{sec:results} we have not performed the latter approximation $t \to \infty$, but take Eq.~\eqref{eq_def_gamma}.

Following the same line of arguments, we can derive an approximate expression for the correlations of 6th Born approximation resulting in a damping $\Gamma_{\mathbf{K}+\mathbf{K}'}F_{\mathbf{K},\mathbf{K'}}^{(i,j),X}$ in the equation of motion for $F_{\mathbf{K},\mathbf{K}'}^{(i,j),X}$. We call this \textit{damped 4th order Born approximation} (4BA-D) where the correlations $S_{\mathbf{K},i}^{(\pm)}$ are not damped by $\Gamma_{\mathbf{K}}$.
\subsection{Time Convolutionless Master Equation}
\label{sec_TCLM}
The TCL master equation is already achieved without taking into account any phonon-assisted correlations explicitly and gives a closed equation already for $P_0(t)$. To derive the TCL master equation we perform similar steps as in the previous section for $S_{\mathbf{K}}^{(\pm)}$. Accordingly, we formally integrate the equations of motion for $S_{\mathbf{K},i}^{(\pm)}$ (Eq.(\ref{eom_Sminus})-(\ref{eom_Splus})) in analogy to Eq.(\ref{eq_4BA_approx}) resulting in
\begin{align*}
	&S_{\mathbf{K},i}^{(\pm)}(t)\approx 
{\gamma}_{\mathbf{K},i,\mathbf{0}}^{(\pm)}(t)\cdot P_0(t) \,.
	\end{align*}
Inserting this in the equation of motion for the coherent polarization already results in a closed form 
\begin{align}
	i\hbar\frac{d}{dt}P_0=\left(E_{1s}-\frac{i\hbar}{\tau}\right)P_0-M_{1s}E(t)-i\hbar\Gamma_0(t)P_0 
	\label{eq_EOM_TCLM}
	\end{align}
where
	$$\Gamma_0(t)=\frac{i}{\hbar}\sum\limits_{\mathbf{K}}\left(\gamma_{\mathbf{K},i,0}^{(-)}(t)+\gamma_{\mathbf{K},i,0}^{(+)}(t)\right)$$
consistent with Eq.(\ref{eq:Gamma}).
In contrast to the former cases, we here do not need to perform the RPA because $P_0$ is the only quantity in 0th order of the exciton-phonon coupling. Because Eq.~(\ref{eq_EOM_TCLM}) is a closed master equation for $P_0$ with a time-dependent dephasing rate, this is equivalent to the TCL master equation usually derived in the theory of open quantum systems~\cite{Breuer02BOOK}.

Due to the simplicity of the equation, we can analytically integrate Eq.~(\ref{eq_EOM_TCLM}) assuming an initial value of $P_0(t=0)$ (which can be achieved e.g. by a delta pulse like excitation). 
\begin{equation}
\label{eq:TCL_P}
	P_0(t)=\Theta (t)e^{-\frac{i}{\hbar}\left(E_{1s}-E_0^{(P)}-\frac{i\hbar}{\gamma}\right)t}e^{\phi(t)}P_0(t=0) \,
\end{equation}
with the Heaviside step function $\Theta$. Here we have defined the polaron shift 
\begin{align}
\label{eq:TCL_EP}
	E_0^{(P)}=\mathcal{P}\sum\limits_{\mathbf{K},i}|g_{\mathbf{K},i}|^2\left[\frac{1+n_{\mathbf{K},i}}{E_{-\mathbf{K}}-E_{1s}+\hbar\omega_{\mathbf{K},i}} \right. \nonumber \\
	\left.	+\frac{n_{\mathbf{K},i}}{E_{\mathbf{K}}-E_{1s}-\hbar\omega_{\mathbf{K},i}}\right]
\end{align}
with the Cauchy principal value $\mathcal{P}$. We also introduced the function $\phi$ as
\begin{align*}
	&\phi(t)=
	\sum\limits_{\mathbf{K},i}|g_{\mathbf{K},i}|^2
	\left[ \int_0^t \frac{e^{-\frac{i}{\hbar}(E_{\mathbf{K}}-E_{1s}-\hbar\omega_{\mathbf{K},i})\tau}-1}{i\hbar\left(E_{\mathbf{K}}-E_{1s}-\hbar\omega_{\mathbf{K},i}\right)}n_{\mathbf{K},i}d\tau \right.
	\\
	+
	&
	\left.\int_0^t \frac{e^{-\frac{i}{\hbar}(E_{-\mathbf{K}}-E_{1s}+\hbar\omega_{\mathbf{K},i})\tau}-1}{i\hbar\left(E_{-\mathbf{K}}-E_{1s}+\hbar\omega_{\mathbf{K},i}\right)}(1+n_{\mathbf{K},i})d\tau\right]-\frac{i}{\hbar}E_0^{(P)}t\,.
\end{align*}
This function $\phi(t)$ contains the information for the lineshape and is defined such that $\mathrm{Im}\left[\phi\right]$ only contains the non-Markovian energy renormalizations due to the inclusion of the polaron shift from Eq.~(\ref{eq:TCL_P}). To understand that the TCL approach results in finite linewidths one may again note that
\begin{align}
\mathrm{Re}\left[\lim\limits_{t\rightarrow\infty}\left(-\frac{i}{\hbar}\gamma^{(+)}_{\mathbf{K},i,0}(t)\right)\right]\propto \delta\left(E_{1s}-E_{-\mathbf{K}}+\hbar\omega_{-\mathbf{K},i}\right)
\label{eq_realGamma}
\end{align}
such that the Markovian limit of Boltzmann scattering rates is included with the $\delta$-function indicating resonant phonon-assisted transitions.

Though this equation of motion is derived by performing one approximation more than the 2BA, it turns out to describe the polarization dynamics particularly well. An example underlining this fact is a two-level system coupled to phonons, for which the exact solution is known \cite{Krummheuer02}, actually coninciding with the solution of the TCL master equation \cite{Richter10}. One thing to note about Eq.~(\ref{eq_EOM_TCLM}) is that it violates the common relation of linear response reading
	$$P(\omega)\propto \chi(\omega)E(\omega)$$
with $\chi$ being the linear susceptibility. This product in frequency space results from the fact that in the time domain the response function only depends on time differences, which is violated by the TCL equation. Because of the TCL formulation, the linear susceptibility is not exactly $\chi(\omega)\propto P(\omega)/E(\omega)$. For linear spectra one usually computes $\chi$ by exciting the system by a delta-pulse such that one has access to the full frequency range. In that case $E(\omega)=E_0=const.$, and the spectral shape of $E(\omega)$ is irrelevant such that $\chi(\omega)\propto P(\omega)$ can be directly calculated via $P(\omega)$ (in general this should hold, if the spectrum of the exciting pulse is broader than the considered spectrum of carriers). Hence for linear spectra the TCL approach works well, while in nonlinear optics more care has to be taken \cite{Richter10}. 

\section{Results}
\label{sec:results}
We now present our results for a MoSe$_2$ monolayer with the parameters given in Appendix~\ref{app_mat_parameters}. We will compare the different theoretical methods for the polarization dynamics and absorption spectra.
\subsection{Polarization dynamics}
\label{sec:results:pol}
\begin{figure}[h!]
\includegraphics[width=\columnwidth]{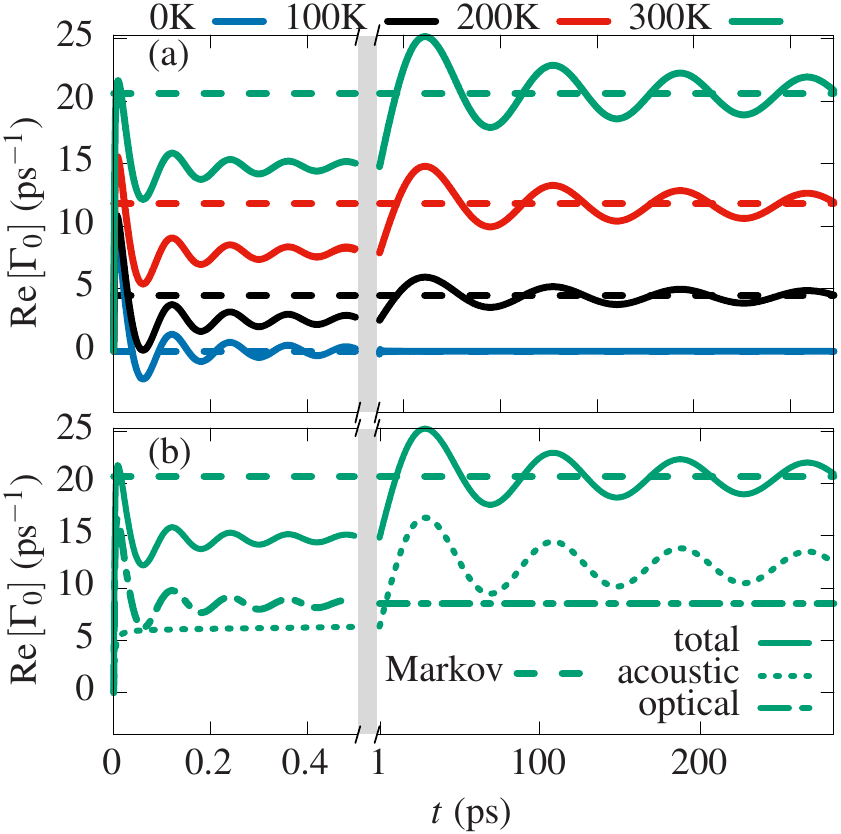}
\caption{\label{fig_realGamma}
Dynamics of the damping rate $\mathrm{Re}\left[\Gamma_0\right]$ according to the TCL formulation (solid lines) as well as its Markovian limit (dashed lines). (a) $\mathrm{Re}\left[\Gamma_0\right]$ for the temperatures $T=0,100,200$ and $300$~K and (b) $\mathrm{Re}\left[\Gamma_0\right]$ split into the contributions from acoustic (dotted) and optical (dashed-dotted) phonons for $T=300$~K.}
\end{figure}

Before calculating the linear absorption spectra, it is instructive to briefly analyze the damping rates and the polarization dynamics as both will be helpful to understand the absorption spectra. In Fig.~\ref{fig_realGamma} we plot the damping rate $\mathrm{Re}\left[\Gamma_0(t)\right]$ obtained in the TCL formulation for four temperatures $T=0,100,200$ and $300$~K. We observe a very fast initial increase of the damping rate followed by an oscillation on short time scales. On longer time scales a larger scale oscillation sets in. For long times all rates approach the Markovian limit of the TCL (cf. Eq.~\eqref{eq_realGamma}), marked by dashed lines. The time to reach the Markovian limit strongly depends on the temperature of the system: While for $0$~K the limit is already reached within $1$~ps, for $300$~K after $300$~ps an oscillation is still visible. This can be traced back to the contribution of the acoustic phonons. This is illustrated in  Fig.~\ref{fig_realGamma}(b), where we plot the different contributions of the acoustic and optical phonons to the damping rate $\mathrm{Re}\left[\Gamma_0(t)\right]$ at $T=300$~K. Here we find that the damping rate attributed to the optical phonons decays rather quickly within about $1$~ps, while the oscillation caused by the acoustic phonons is rather long-lived and results in the oscillation on the longer scale. This explains the temperature dependence of the relaxation to the Markovian limit in Fig.~\ref{fig_realGamma}(a) when considering the two different dispersion relations. For optical phonons the constant dispersion relation of $\hbar\omega_{\mathrm{opt}}=34.4$~meV leads to a weak dependence on temperature because optical phonon absorption only becomes important for thermal energies above $\hbar\omega_{\mathrm{opt}}$. The acoustic phonons have a linear dispersion relation such that they show a stronger dependence on temperature and acoustic phonon absorption becomes important at elevated temperatures as seen in Fig.~\ref{fig_realGamma}(b). The separation of time scales for optical and acoustic phonons also relate to the two oscillation frequencies visible in Fig.~\ref{fig_realGamma}(b). On the time scale up to 1~ps a fast oscillation with the frequency $\omega_{\mathrm{opt}}$ of the optical phonons is seen. A much smaller frequency dictated by acoustic phonons  dominates the time scale after 1~ps and is determined by details of the exciton and phonon modelling.

\begin{figure}[t!]
	\includegraphics[width=\columnwidth]{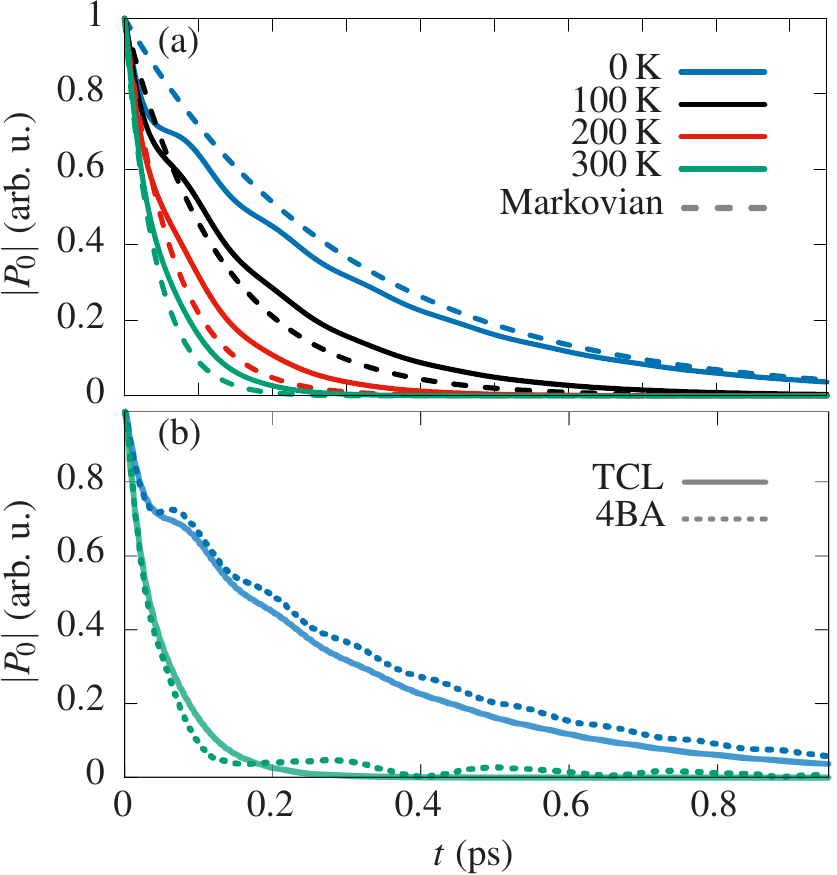}
	\caption{Dynamics of the polarization $|P_0(t)|$. (a) Calculation using the TCL approach (solid lines) and Markovian rates (dashed lines) for $T=0,100,200,$ and $300$~K . (b) Calculation with the TCL approach (solid) and 4BA (dotted) for $T=0$~K and $T=300$~K. }
	\label{fig_poldyn}
\end{figure}

We now look at the polarization dynamics shown in Fig.~\ref{fig_poldyn}(a), where we plot $|P_0(t)|$ for different temperatures for the TCL approach (solid lines) and the Markovian approach (dashed lines), where in both cases a radiative dephasing time of $\tau=300$~fs has been added. The polarization decays quickly within the first $0.5$~ps. For the Markovian approach the polarization goes exponentially to zero. The TCL approach shows an exponential decay with small oscillations which are superimposed. Because the polarization has decayed within the first ps, the long-term oscillation of the damping rate due to the acoustic phonons in Fig.~\ref{fig_realGamma} does not contribute to the dynamics anymore. We further note that the polarization has decayed long before $\mathrm{Re}\left[\Gamma_0(t)\right]$ has relaxed to the Markovian limit, which shows that due to the strong exciton-phonon interaction the relevant time scale for the polarization is much smaller than the time scale in which the Markovian limit is reached such that strong non-Markovian features can be expected. 

We further compare the results of the TCL to the quantum kinetic approach in 4th Born approximation. To this end, we plot in Fig.~\ref{fig_poldyn}(b) the dynamics of the polarization $|P_0(t)|$ for the two approaches at $T=0$~K and $T=300$~K. For $0$~K one observes very comparable dynamics apart from the fact that the TCL master equation predicts a slightly stronger dephasing. At $300$~K the TCL approach shows approximately an exponential decay, while we find a slightly quicker decay using the Born approximation in the first $0.1$~ps. More interestingly, we see that after the initial decay some oscillations remain in the polarization in the case of 4BA. We will come back to this point later. 

\subsection{Absorption spectra}
\label{sec:results:abs}
\begin{figure}[h]
	\includegraphics[width=\columnwidth]{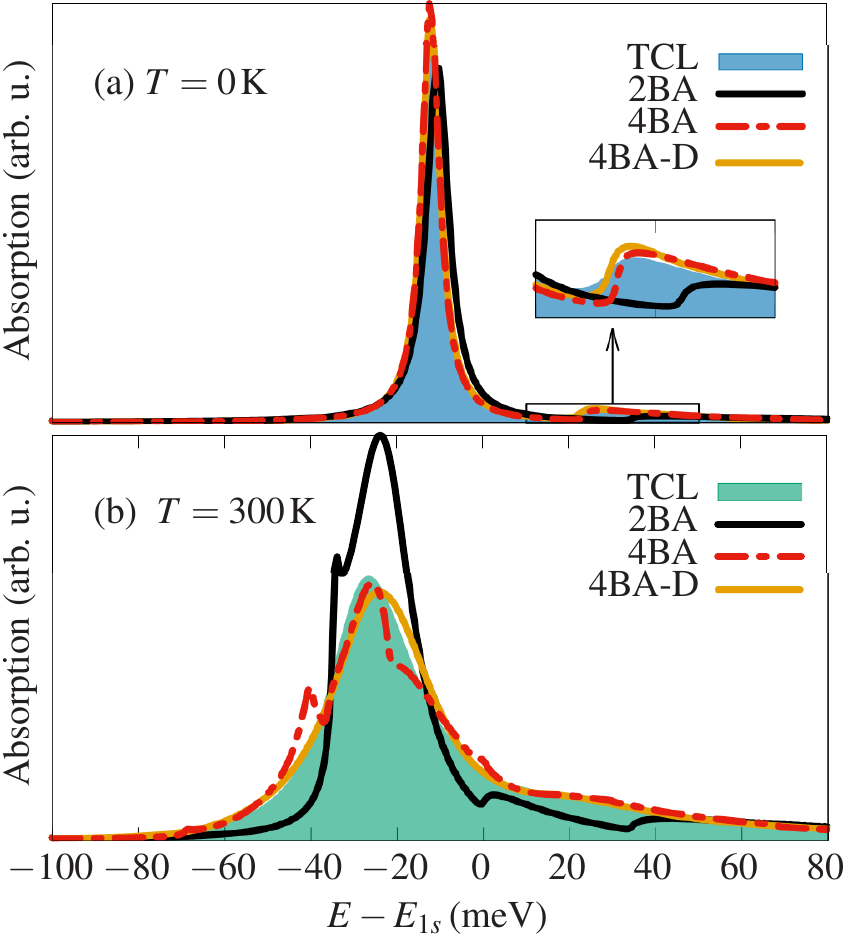}
	\caption{Calculated absorption spectrum of a MoS$_2$ monolayer using different methods for (a) $T=0\,$K and (b) $T=300\,$K.}
	\label{fig_abs_compare}
\end{figure}

Now we turn to the absorption spectra. Accordingly, we show the absorption spectra for the different approximations 2BA (black line), 4BA (dashed red line), 4BA-D (orange solid line) and TCL (shaded area) for $T=0$\,K in Fig.~\ref{fig_abs_compare}(a) and for $T=300$~K in Fig.~\ref{fig_abs_compare}(b)

Without phonons, we would obtain a Lorentzian peak at $E-E_{1s}=0$ with the broadening given by the radiative decay. When the exciton-phonon interaction is active, for $T=0$~K we find also a single peak for all calculation methods with a rather similar shape. The peak is at  $E-E_{1s}= -12$~meV due to a strong polaron shift dominated by optical phonons with  a comparable linewidth for all four methods. In the TCL approach the linewidth is a bit larger than in 4BA due to the stronger dephasing seen in Fig.~\ref{fig_poldyn}(b). In addition, we see a phonon sideband originating from emission of optical phonons at the positive energy tail (amplified in the inset). Here, we find a small difference in the four methods: While in 4BA and TCL the phonon sideband starts at $E-E_{1s}\approx 24\,$meV, which is about one optical phonon energy above the polaron, for the 2BA it starts at $E-E_{1s}\approx 34\,$meV for 2BA, which is one optical phonon energy above $E_{1s}$. This shows that the 2BA does not correctly take into account the ground state of the system since the phonon sideband lies one phonon energy relative to the unperturbed ground state. This is apparent in the equations of motion in 2BA since the free-oscillating part on the right hand side of Eq.~(\ref{eom_Sminus}) and Eq.~(\ref{eom_Splus}) contains the unperturbed energies whereas the RPA treatment of the 4BA shows that higher correlations lead to a renormalization of these single-particle energies in Eq.~(\ref{eq_4BA_approx}). It is a general feature of the correlation expansion that processes in highest order of the respective expansion are not correctly renormalized. We will show in Sec.~\ref{sec:results:comparison} analytically that the TCL approach is free of this problem even though the derivation of the TCL starts with the equations of motion of 2BA. Comparing 4BA and 4BA-D one only observes minor differences indicating that for $0$~K 6th order contributions from the exciton-phonon interaction are not important.

We now turn to the higher-temperature case and consider the absorption spectra at $T=300$~K in Fig.~\ref{fig_abs_compare}(b). Here, the different methods give rather different results. The only aspect which all methods have in common is an increased polaron shift of about $27\,$meV, while the linewidth and shape differ greatly. In 2BA we find a much smaller linewidth compared to the other methods. From this we can conclude that the broadening of the spectrum is a result of a damping of one-phonon-assisted correlations by two-phonon-assisted correlations as included in the other methods. When comparing the phonon sidebands at the positive energy tail we also find that 2BA gives pronounced features, while in 4BA and TCL these are rather smeared out. In addition, we find a sharp sidepeak in 2BA around $-35$~meV. In 4BA there are similar sharp features from phonon sidebands in the spectrum.  In contrast, the 4BA-D method and the TCL method both show a smooth line broadening and all sharp features are gone. We note that 4BA-D and TCL are rather similar. To understand the difference in the sharp spectra and the smooth line, i.e. comparing in particular 4BA and 4BA-D/TCL we go back to the polarization dynamics as seen in Fig.~\ref{fig_poldyn}(b). These sharp features seen in 4BA originate from the long-term oscillation in the polarization dynamics, which is rather unexpected. Reminding that in all calculations a radiative dephasing time of $\tau=300$~fs is included, almost no polarization should be left in the system after $1$~ps just due to this dephasing. This means that the exciton-phonon correlations in 4BA are so strong that they overcompensate the radiative dephasing. We remind that 4BA neglects 3-phonon-assisted correlations, which in principle would damp the two-phonon-assisted correlations. The latter damping of correlation is accounted for in 4BA-D, which then shows a smooth behavior. This indicates that the sharp peaks in 4BA are an artifact caused by truncating the hierarchy of equation of motions too early. 

Interestingly, the TCL gives rather similar results to the 4BA-D. Only the polaron shift is slightly larger in TCL than in 4BA-D compared to TCL, but this difference is barely visible. All essential features like lineshape, polaron shift and sidebands are qualitatively correctly described by the TCL approach underlining the predictive power of this approach. We emphasize how remarkable this is because the computational complexity of 4BA-D is orders of magnitudes higher than for the TCL master equation.

\begin{figure}[t]
	\includegraphics[width=\columnwidth]{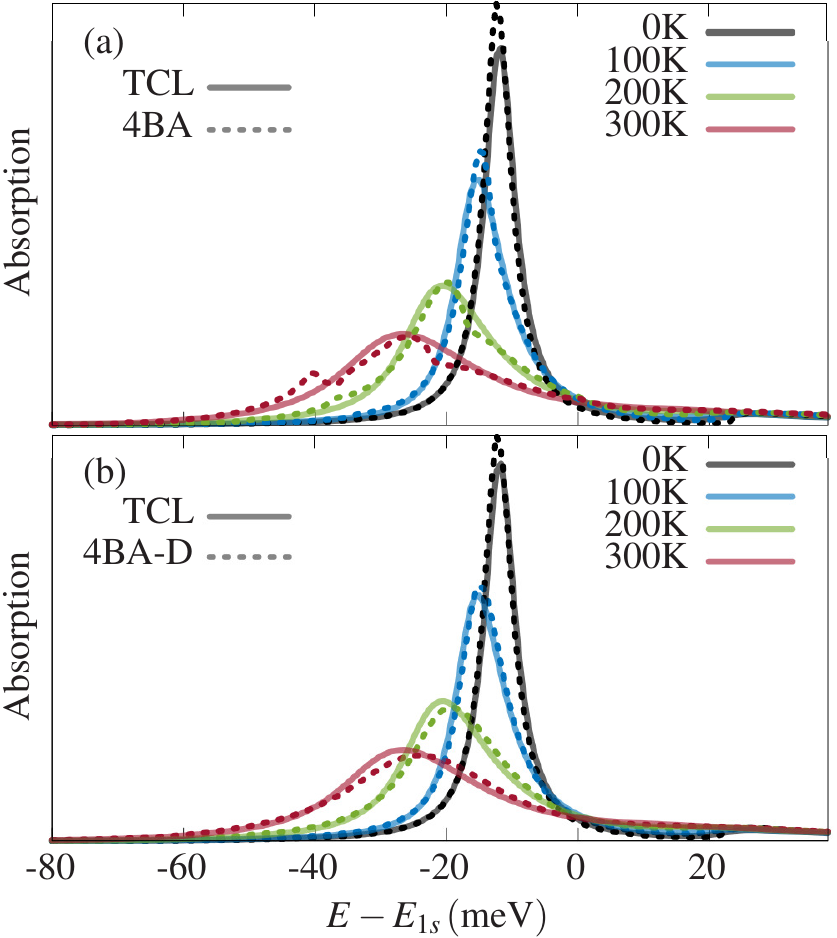}
	\caption{Absorption spectrum for a temperature range from $T=0\,$K to $T=300\,$K (a) Comparison of TCL (solid) to 4BA (dotted) and (b) comparison of TCL (solid) to 4BA-D (dotted).}
	\label{fig_abs_temp}
\end{figure}

Furthermore, we show that the TCL method compares rather well with the 4th Born approximation in a wide range of temperatures as shown in Fig.~\ref{fig_abs_temp}, where we compare 4BA to TCL in Fig.~\ref{fig_abs_temp}(a) and 4BA-D to TCL in Fig.~\ref{fig_abs_temp}(b) for the full temperature range from $T=0\,$K to $T=300\,$K. We again find a very good agreement of both methods with TCL and that all sharp spectral features of 4BA are smoothed out by damping in 4BA-D. While the overall agreement of 4BA-D and TCL in Fig.~\ref{fig_abs_temp} is very good, a slightly larger polaron shift is predicted by TCL. 

\begin{figure}[]
	\includegraphics[width=\columnwidth]{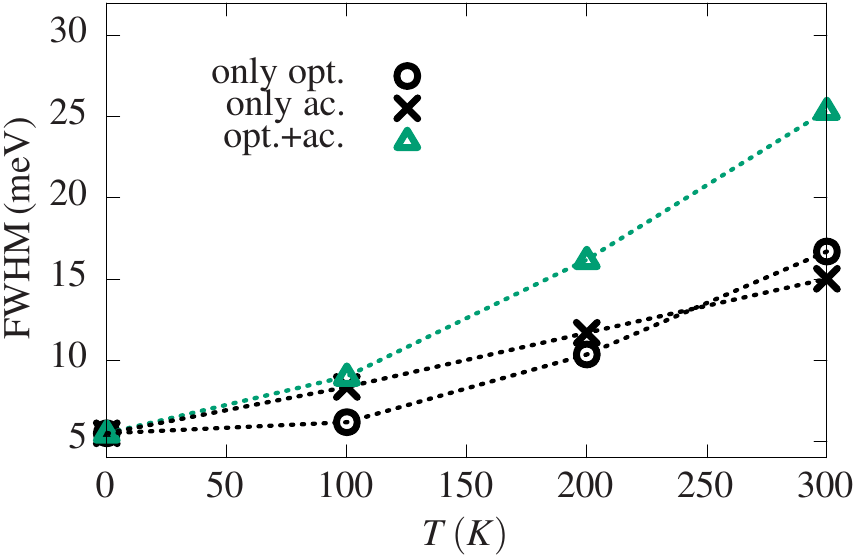}
	\caption{FWHM as a function of temperature calculated by TCL considering only optical (circles), only acoustic (crosses) and both phonon branches (triangles). Dotted lines are added as guide to the eyes.}
	\label{fig_FWHM}
\end{figure}

We emphasize that the lineshape especially for elevated temperatures does not follow a Lorentzian shape, but is asymmetrically broadened. To quantify the phononic impact on the line broadening, we therefore plot the full width at half maximum (FWHM) of the absorption spectrum as a function of temperature in Fig.~\ref{fig_FWHM} for the TCL approach. We checked that the FWHM for 4BA and 4BA-D are similar (cf. Fig.~\ref{fig_abs_temp}). We find a non-linear increase of the FWHM from 5~meV to 25~meV in the considered temperature range. The value at 0~K essentially results from the radiative lifetime $\tau$ (cf. Eq.~(\ref{eq:QK_P})). 
To discriminate the influence of the different phonon branches, we additionally plot the FWHM considering only optical (circles) or only acoustic (crosses) phonons. When only acoustic phonons are taken into account, a strictly linear increase is found. For optical phonons, the FWHM increases non-linearly following from the Bose distribution $(\exp(\hbar\omega_{\mathrm{opt}}/(k_BT))-1)^{-1}$. These dependencies for acoustic and optical phonons are in agreement with the theoretical and experimental results in Ref.~\cite{Selig16}.

Measurements of the absorption spectra of a TMDC monolayer have been reported in Refs.~\cite{Christiansen17,Shree18} and we briefly want to compare our findings with the measured data. First of all the strong temperature dependence and strong line broadening shown in 4BA/TCL is consistent with these experimental findings. The measured spectra were always very smooth and no sharp features as found in our calculations were observed. This underlines our interpretation that even the 4BA truncates the equations of motion too early and that high-order correlations are important. One aspect, which might result in an additional inhomogeneous broadening is disorder, which we did not take into account. The TCL and 4BA-D show strong polaron shifts comparable with previous theoretical calculations performed within a self-consistent calculation of the Markovian dephasing which was then reinserted in 2BA \cite{Selig16,Christiansen17}. This is reasonable because the polaron shift is well described already by 2BA. While our calculated FWHM show the same trend as in Ref.~\cite{Christiansen17}, the quantiative values there are higher than our predicted values which may be linked to the specific model parameters of the exciton-phonon interaction. Furthermore, for comparison with experiments disorder may influence the quantitative values. Our calculations also underline that it is important to include damping of phonon-assisted correlations in order to deal with unphysical sharp replica which is done by the Markovian dephasing rate in Refs.~\cite{Selig16,Christiansen17}. The presented calculations therefore give important insights on the effect of higher-order correlations on the line width broadening as function of temperature. 
Additionally it is seen that a Markovian treatment of dephasing rates may overestimate the linewidth at higher temperatures due to the discussion of Fig.~\ref{fig_realGamma}.

\subsection{Analytical comparison of TCL and BA}
\label{sec:results:comparison}
Having seen that the TCL approach gives surprisingly good results, we perform an analytical analysis of the solution of TCL and n-th order Born approximation (nBA) of the quantum kinetic approach. For sake of simplicity we consider the $T=0$~K case with $n_{\mathbf{K},i}=0$. Additionally we drop the $i$-index of all phononic quantities. Setting $-M_{1s}E_0(\omega)=\mathcal{E}$ as a constant due to the excitation with a delta pulse, in the TCL approach the polarization (Eq.~\eqref{eq:TCL_P}) can be easily transformed to frequency space to 
	\begin{widetext}
	\begin{align}\label{eq_pol_TCLM}
	P_{\mathrm{TCL}}(\omega)=&\frac{1}{\hbar\omega-E_{1s}+i\gamma_0}\left(\mathcal{E}+\sum\limits_{\mathbf{K}} \frac{|g_{\mathbf{K}}|^2}{E_{1s}-E_{\mathbf{-K}}-\hbar\omega_{\mathbf{K}}}\left[P_{TCL}(\omega)-P_{TCL}\left(\omega+\frac{E_{1s}-E_{\mathbf{-K}}-\hbar\omega_{\mathbf{K}}}{\hbar}\right)\right]\right) \,\\
	=&\frac{\mathcal{E}}{\hbar\omega-E_0^{(P)}+i\gamma_0} -\sum\limits_{\mathbf{K}}\frac{|g_{\mathbf{K}}|^2(\hbar\omega-E_0^{(P)}+i\gamma_0)^{-1}}{E_{1s}-E_{\mathbf{-K}}-\hbar\omega_{\mathbf{K}}}
	  P_{TCL}\left(\omega+\frac{E_{1s}-E_{\mathbf{-K}}-\hbar\omega_{\mathbf{K}}}{\hbar}\right) \,. \nonumber	
	\end{align}
\end{widetext}
In the second line we rewrote the equation using the polaron energy $E_0^{(P)}$ from Eq.~\eqref{eq:TCL_EP} where no singularity occurs because of $T=0~$K such that one does not need to take care of the principal value. The latter formulation allows us to see that the fundamental resonance of the TCL approach is exactly at the polaron-shifted transition energy (cf. the first term), while the second term describes a self-consistent formulation of phonon-assisted processes resulting in the lineshape. This self-consistent formulation resembles the description of retarded Green functions with an exciton self energy in frequency space \cite{Shree18}, nevertheless the one given here results in a closed equation of motion and can be analytically integrated.

In order to compare the TCL equation with the nBA we now expand the polarization $P(\omega)$ from Eq.~(\ref{eq_pol_TCLM}) in orders $N$ of the phonon coupling matrix element ${g_{\mathbf{K}}}$ using
$$
P_{\mathrm{TCL}}(\omega)= \sum_N P_{\mathrm{TCL}}^{(N)}
$$
The two lowest orders read
\begin{align*}
	P_{\mathrm{TCL}}^{(0)}=&\frac{\mathcal{E}}{\hbar\tilde{\omega}-E_{1s}}\\
	P_{\mathrm{TCL}}^{(2)}=
	&\frac{\mathcal{E}}{\left(\hbar\tilde{\omega}-E_{1s}\right)^2}\sum\limits_{\mathbf{K}}\frac{|g_{\mathbf{K}}|^2}{\hbar\tilde{\omega}-E_{\mathbf{-K}}-\hbar\omega_{\mathbf{K}}} \, , 
\end{align*}
where we abbreviated $\hbar\tilde{\omega}=\hbar\omega+i\gamma_0$. 
 
Since the nBA is exact up to n-th order the comparison of $P_{\mathrm{TCL}}^{(N)}$ with the nBA will give an insight into the accuracy of the TCL approach. For the nBA we also Fourier transform  the equations of motion from Sec.~\ref{sec:QK_BA}. For the 0-th order we consider Eq.~\eqref{eq:QK_P} and set $S_{\mathbf{K},i}^{(\pm)}$ to zero. For the 2nd order (2BA) we explicitly take $S_{\mathbf{K},i}^{(\pm)}$ into account and set the next order, namely $F_{\mathbf{K},\mathbf{K'}}^{(i,j)X}$, to zero. Then we perform a Fourier transform of Eq.~\eqref{eom_Sminus} and Eq.~\eqref{eom_Splus} and plug this in Eq.~\eqref{eq:QK_P}. For the comparison, we also only give the results up to the n-th order of the phonon coupling matrix element ${g_{\mathbf{K}}}$. Then the two lowest orders read
\begin{align*}
	P_{\textrm{BA}}^{(0)}=&\frac{\mathcal{E}}{\hbar{\tilde{\omega}}-E_{1s}}\\
	P_{\textrm{BA}}^{(2)}=
	& \frac{\mathcal{E}}{\left(\hbar{\tilde{\omega}}-E_{1s}\right)^2}\sum\limits_{\mathbf{K}}\frac{|g_{\mathbf{K}}|^2}{\hbar{\omega}-E_{-\mathbf{K}}-\hbar\omega_{\mathbf{K}}}.
\end{align*}
Comparing now $P^{(2)}_{\mathrm{TCL}}$ to $P_{\mathrm{BA}}^{(2)}$, we find that the solutions agree except of a difference of $\hbar\omega\rightarrow \hbar\tilde{\omega}$ in the denominator within the sum. The latter results in a radiative broadening of the phonon-assisted processes due to the fact that in the self-consistent formulation of the TCL approach everything is expanded with respect to the optically active resonance which is broadened by radiative decay. We nevertheless note that a change of the radiative decay did not change the spectra too much (not shown here) and at high temperatures the phonon-induced broadenings are dominating. Having in mind that especially for high temperatures the spectra in TCL and 2BA did not match at all, this means that the linear response at high temperatures is dominated by higher-order processes. 

For the 4-th order we obtain using the same procedures
\begin{widetext}
	\begin{align*}
	P_{\mathrm{TCL}}^{(4)}=
&\sum\limits_{\mathbf{K},\mathbf{Q}} \frac{|g_{\mathbf{K}}|^2|g_{\mathbf{Q}}|^2\mathcal{E}}{(\hbar\tilde{\omega}-E_{1s})^3(\hbar\tilde{\omega}-E_{-\mathbf{K}}-\hbar\omega_{\mathbf{K}})(\hbar\tilde{\omega}-E_{-\mathbf{Q}}-\hbar\omega_{\mathbf{Q}})}\\
&+\sum\limits_{\mathbf{K},\mathbf{Q}} \frac{|g_{\mathbf{K}}|^2|g_{\mathbf{Q}}|^2\mathcal{E}}{(\hbar\tilde{\omega}-E_{1s})^2(\hbar\tilde{\omega}-E_{-\mathbf{K}}-\hbar\omega_{\mathbf{K}})^2}\frac{2\hbar\tilde{\omega}-E_{-\mathbf{K}}-\hbar\omega_{\mathbf{K}}-E_{-\mathbf{Q}}-\hbar\omega_{\mathbf{Q}}}{(\hbar\tilde{\omega}-E_{-\mathbf{Q}}-\hbar\omega_{\mathbf{Q}})(\hbar\tilde{\omega}+E_{1s}-E_{-\mathbf{K}}-E_{-\mathbf{Q}}-\hbar\omega_{\mathbf{K}}-\hbar\omega_{\mathbf{Q}})}\\
	P_{\textrm{BA}}^{(4)}=&
\sum\limits_{\mathbf{K},\mathbf{Q}} \frac{|g_{\mathbf{K}}|^2|g_{\mathbf{Q}}|^2\mathcal{E}}{(\hbar\tilde{\omega}-E_{1s})^3(\hbar{\omega}-E_{-\mathbf{K}}-\hbar\omega_{\mathbf{K}})(\hbar{\omega}-E_{-\mathbf{Q}}-\hbar\omega_{\mathbf{Q}})}\\
&+\sum\limits_{\mathbf{K},\mathbf{Q}} \frac{|g_{\mathbf{K}}|^2|g_{\mathbf{Q}}|^2\mathcal{E}}{(\hbar\tilde{\omega}-E_{1s})^2(\hbar{\omega}-E_{-\mathbf{K}}-\hbar\omega_{\mathbf{K}})^2}\frac{2\hbar{\omega}-E_{-\mathbf{K}}-\hbar\omega_{\mathbf{K}}-E_{-\mathbf{Q}}-\hbar\omega_{\mathbf{Q}}}{(\hbar{\omega}-E_{-\mathbf{Q}}-\hbar\omega_{\mathbf{Q}})(\hbar{\omega}-E_{-\mathbf{K}-\mathbf{Q}}-\hbar\omega_{\mathbf{K}}-\hbar\omega_{\mathbf{Q}})}.
	\end{align*}

\end{widetext}

Now again, we find a surprisingly good agreement.  Again all phonon-processes are radiatively broadened in TCL compared to 4BA as in the case of 2BA. However the major difference between the terms can be found in the denominator of the last term, where the energy denominator of the two approaches read
	\begin{align}
	\textrm{TCL: }&\hbar\tilde{\omega}+E_{1s}-E_{-\mathbf{K}}-E_{-\mathbf{Q}}-\hbar\omega_{\mathbf{K}}-\hbar\omega_{\mathbf{Q}} 	
	\label{eq_TCLM_energy_approx}
\\
	\textrm{4BA: }&\hbar\omega-E_{-\mathbf{K}-\mathbf{Q}}-\hbar\omega_{\mathbf{K}}-\hbar\omega_{\mathbf{Q}}.  \nonumber
	\end{align}
This difference can be interpreted as follows: while in 4BA a true 2-phonon process occurs where two phonons together carry the momentum transfer $\textbf{K+Q}$, in the TCL approach the total momentum is transferred in 2 separate processes of momenta $\textbf{K}$ and $\textbf{Q}$. The emission of two separate phonons in the TCL acts very similar to a 2-phonon process, because it causes resonances where the denominator of $P$ is $0$. This observation also explains why the TCL gives the exact result for a two-level system coupled to phonons (when radiative dephasing is not considered) \cite{Richter10} since then the expressions in ~(\ref{eq_TCLM_energy_approx}) are identical with $E_{\mathbf{K}}\rightarrow E_{1s}\forall \mathbf{K}$. Another example where the expressions in ~(\ref{eq_TCLM_energy_approx}) are identical is the case of linear dispersion with $E_{\mathbf{K}}-E_{1s}\propto K$ in one dimension. For this case the problem of intraband relaxation by optical phonons has been analytically solved in Ref.~\cite{Meden95} and a comparison to quantum kinetic models similar to the presented study has been carried out in Ref.~\cite{Fricke97}. In Refs.~\cite{Meden95,Fricke97} an exact equation of motion for the one-particle density matrix has been derived which coincides with the equation of motion in TCL consistent with our prediction by~(\ref{eq_TCLM_energy_approx}).

It is interesting to note that the first term of $P_{\mathrm{BA}}^{(4)}$ stems from 2BA ($P_{\mathrm{2BA}}^{(4)}$) and can be written as
	\begin{align*}
	P_{\textrm{2BA}}^{(4)}=&
\left(P_{\mathrm{BA}}^{(2)}\right)^2\frac{\hbar\tilde{\omega}-E_{1s}}{\mathcal{E}}.
	\end{align*}
This shows that 2BA contains contributions of 4th order, but only describes independent phonon processes because $P_{\mathrm{2BA}}^{(4)}$ is a bare product of two one-phonon processes.

This analytical comparison shows that TCL is a very accurate method to calculate absorption spectra, because it is exact up to second order in the exciton-phonon coupling (apart from radiatively broadened phonon-assisted transitions) and it can - unlike the bare correlation expansion - treat phonon-assisted transitions of arbitrary number of phonons approximately due to its self-consistent structure in the frequency domain. The radiative broadening affecting the phonon-assisted transition also explains why the phonon sideband in the inset of Fig.~\ref{fig_abs_compare}(a) in TCL is a bit broadened compared to 4BA. This broadening is however very small and underlines that the radiative broadening does not change the spectra too much.\\
We note that the TCL method is easily extendable to W-based TMDCs where intervalley processes and dark excitons play an important role \cite{Selig16,Malic18}. In the W-based TMDCs multiple excitonic valleys are linked by phonons, which can be naturally included in the TCL approach.\\
Analogously higher excited excitonic states above the $1s$ state can be included in the description by accounting for multiple excitonic bands and their coupling by phonons in the Hamiltonian in Eq.~(\ref{eq:Hamiltonian}). 

\subsection{Weak Exciton-Phonon Interaction}
\label{sec:results:weak}
\label{sec_weak}
	\begin{figure}[ht]
	\includegraphics[width=\columnwidth]{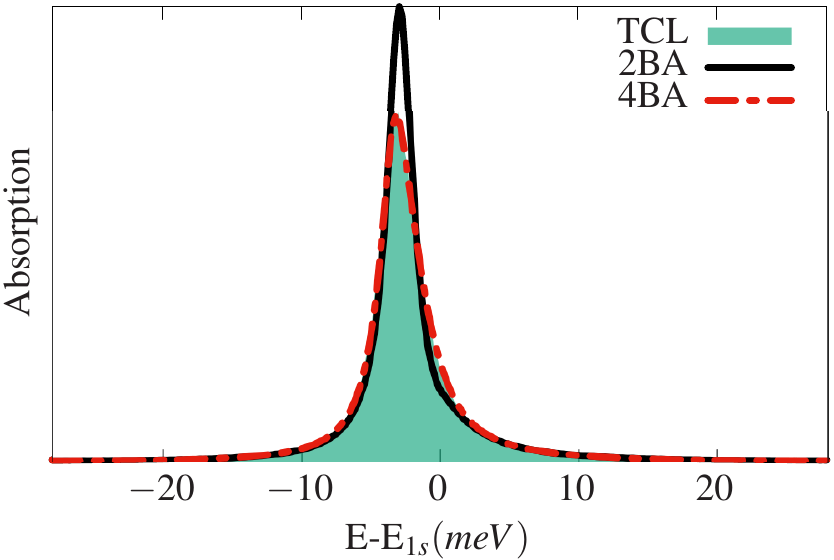}
	\caption{Absorption spectra for scaled exciton-phonon interaction with $0.1 \cdot |g_{\mathbf{K},i}|^2$ using different methods for $T=300$~K.}
	\label{app_fig_weak_compare}
	\end{figure}
For the case of strong electron-phonon interaction, we have seen that the BA methods do not give reasonable results when using the lowest orders only at high temperatures. For semiconductors with a rather weak electron-phonon interaction, the BA method has been widely used and also when calculating other quantities like transport. Therefore, is it interesting to study, whether for low electron-phonon coupling strength, the different methods agree well. For this, we reduce the exciton-phonon matrix elements by a factor of $\sqrt{10}$
$$
|g_{\mathbf{K},i}|^2\rightarrow 0.1 \cdot |g_{\mathbf{K},i}|^2 \,.
$$
The resulting absorption spectra for 4BA, 2BA and TCL for $T=300\,$K are shown in Fig.~\ref{app_fig_weak_compare}. We here only consider the high temperature case, because there we have found the largest differences. 

For the low exciton-phonon interaction we find a very good agreement between the TCL and the 4BA, while the 2BA gives a similar overall shape with slightly different polaron shift and line broadening. 

\section{Conclusion}
\label{sec:concl}
In this paper we have derived and compared different theoretical models to calculate the linear response to a light field of a semiconductor under the influence of exciton-phonon interaction where we reduced our attention to the $1s$ exciton of a TMDC monolayer. We have shown that in TMDCs the exciton-phonon coupling is so strong that low-order treatments of the interaction lead to poor results, especially at high temperatures. Even the correlation expansion up to 4th order (4BA) still shows artifacts only resolved by adding damping functions from 6th order (4BA-D). The central aspect of this work is the identification of a time convolutionless formulation (TCL) as a very easy and extremely precise method to calculate absorption spectra. We showed that the TCL approach gives excellent results in the whole temperature range when compared to 4BA-D and furthermore that the TCL is exact up to second order in the coupling (if spontaneous decay is not dominating) and approximately accounts for processes of arbitrarily many phonons. We therefore believe that this method is of great value for systems where strong exciton-phonon interaction is present as, e.g., in the recently very popular monolayer materials.

\section*{Acknowledgements}
 F.~L. and D.~E.~R. acknowledge financial support by the Deutsche Forschungsgemeinschaft (DFG) through the project
406251889 (RE 4183/2-1). 

\appendix 
\section{Details of the model and material parameters}
\label{app_mat_parameters}
We describe our TMDC monolayer in a two band model consisting of a conduction band and a valence band. The corresponding electron and hole creation operators are $\cre{c}_{\mathbf{k}\uparrow}$ and $\cre{d}_{\mathbf{k}\downarrow}$, respectively. We restrict our attention to one spin component leading to the A exciton at the $K$-valley. For the description of the exciton parabola, we make use of the exciton creation operators, which are derived via 
$$
	\cre{P}_{\mathbf{K}}=\sum\limits_{\mathbf{k}} \phi_{1s}\left(\mathbf{k}-\frac{m_e}{M}\mathbf{K}\right)\cre{c}_{\mathbf{k}\uparrow}\cre{d}_{\mathbf{K-k}\downarrow}
$$ 
creating an exciton with center-of-mass momentum $\mathbf{K}$ at the $K$-valley with energy $E_{\mathbf{K}}=E_{1s}+\frac{\hbar^2}{2(m_e+m_h)}K^2$ using the effective masses of electrons ($m_e$) and holes ($m_h$) and bound state energy $E_{1s}$. $\phi_{1s}(\mathbf{k})$ is the $1s$-exciton wave function calculated via the Wannier equation.

From Ref.~\cite{Rasmussen15} we extract the effective masses of electrons $m_e$ and holes $m_h$ used here as $m_e=0.49m_0$ and $m_h=0.61m_0$ where $m_0$ is the free electron mass. The Coulomb potential forming excitons is approximated as the Keldysh potential \cite{Cudazzo11,Berghaeuser14}
	$$V_{\mathbf{q}}=\frac{e^2}{2\epsilon_0\epsilon_s A}\frac{1}{q(1+r_0q)}$$
with the normalization area $A$, the vacuum permittivity $\epsilon_0$, the unit charge $e$, the effective dielectric constant $\epsilon_s=(\epsilon_1+\epsilon_2)/2=(1+3.9)/2$ approximating the screening of a monolayer on top of a $\mathrm{SiO_2}$ substrate ($\epsilon_2=3.9$) with air ($\epsilon_1=1$) above and the screening length $r_0=d\epsilon_{\perp}/\epsilon_s$ with the dielectric screening $\epsilon_{\perp}=15.3$ of $\mathrm{MoSe_2}$ \cite{Malic18} and the distance $d=0.334\,$nm \cite{Rasmussen15} between selenium atoms approximating the thickness of the sample.

 The exciton-phonon interaction is derived from the electron-phonon and hole-phonon interaction and can exactly be written in the given form in the case of linear response since \cite{Axt94}
	\begin{align}
	&\cre{c}_1\hat{c}_2=\sum\limits_{3} \cre{c}_1\cre{d}_3\hat{d}_3\hat{c}_2+\mathcal{O}\left(E^4\right)\\
	&\cre{d}_1\hat{d}_2=\sum\limits_{3} \cre{d}_1\cre{c}_3\hat{c}_3\hat{d}_2+\mathcal{O}\left(E^4\right)
	\end{align}
for a system being in the ground state before optical excitation.

The exciton-phonon interaction is then given by
\begin{align}
	g_{\mathbf{Q},i}=g_{e,i}(\mathbf{Q})F_{1s}(\mu_h\mathbf{Q})-g_{h,i}(\mathbf{Q})F_{1s}(\mu_e\mathbf{Q}) 
	\label{eq_Xph_coupling}
\end{align}
with the electron/hole couplings $g_{e/h,i}(\mathbf{Q})$ of branch $i$ and the exciton form factor
	$$F_{1s}(\mathbf{Q})=\sum\limits_{\mathbf{k}}\phi_{1s}^*(\mathbf{k})\phi_{1s}(\mathbf{k}-\mathbf{Q})\,. $$

\begin{table}[h!]
\caption{Phonon parameters. Originating from \cite{Jin14} and calculated as described in the text.}
\label{table_phonon_params}
\begin{tabular}{c|c|c}
\hline
Parameter & Symbol & Value\\
\hline\hline
Unit cell density & $\rho$ & $4.26\times 10^{-7}\,\mathrm{\frac{g}{cm^2}}$\\
Opt. phonon energy & $\hbar\omega_{\mathrm{opt}}$ & $34.4\,$meV  \\
Sound velocity & $c_{s}$ & $4.1~\mathrm{\frac{nm}{ps}}$\\
Opt. deformation potential (electrons) & $D_{0,\mathrm{opt}}^e$ & $52\,\mathrm{\frac{eV}{nm}} $\\
Opt. deformation potential (holes) & $D_{0,\mathrm{opt}}^h$ & $-49\,\mathrm{\frac{eV}{nm}} $\\
Ac. deformation potential (electrons) & $D_{1\mathrm{ac}}^e$ & $2.40\,$eV\\
Ac. deformation potential (holes) & $D_{1\mathrm{ac}}^h$ & $-1.98\,$eV\\
\hline
\end{tabular}
\end{table}
	
We describe the carrier-phonon interaction by a deformation potential approximation in which
	\begin{align}
	&g_{e/h,i}(\mathbf{q})=\sqrt{\frac{\hbar}{2A\rho\omega_{i,\mathbf{q}}}}\Delta V_{\mathbf{q},i}^{e/h}\\
	&\Delta V_{\mathbf{q},i}^{e/h}\approx D_{0,i}^{e/h}+D_{1,i}^{e/h}{q}
	\end{align}
with the unit cell density $\rho$ (taken as $(2M_{\mathrm{Se}}+M_{\mathrm{Mo}})\left(\frac{\sqrt{3}}{2}a_0^2\right)^{-1}$ with the atomic masses $M_{\mathrm{Se}},M_{\mathrm{Mo}}$ of selenium and molybdenum and the lattice constant $a_0=0.332\,$nm \cite{Rasmussen15}), the normalization volume $A$, the phonon frequency $\omega_{i,\mathbf{q}}$ of branch $i$ and the bandshift $\Delta V_{\mathbf{q},i}^{e/h}$ of band $e/h$ associated with the mode $(i,\mathbf{q})$. In the approximation the bandshift $\Delta V$ is given by a taylor series in $\mathbf{q}$ whose constants $D_0, D_1$ can be determined by DFT calculations \cite{Kaasbjerg12,Kaasbjerg13,Li13,Jin14}. 
 We here follow the description of Refs.~\cite{Li13,Jin14} where the influence of all phonon branches are approximated by only one optical and one acoustic branch. Therein one approximates all optical phonons around the phononic $\Gamma$ point (LO,TO,A') as one optical phonon with the mean frequency of LO,TO,A' and the same is done for LA,TA phonons. This results in one averaged deformation potential parameter for each band for optical and acoustic phonons, respectively, where optical phonons are treated in 0th order and acoustic ones in 1st order of $q$ at the phononic $\Gamma$ point which are the respective lowest orders. The parameters taken from Ref.~\cite{Jin14} are listed in table \ref{table_phonon_params}. Some care has to be taken when considering acoustic phonons \cite{Kaasbjerg13} because in the effective deformation potential constant from Ref.~\cite{Jin14} the piezoelectric coupling is implicitly included. Piezoelectric coupling is ineffective for excitons being neutral particles and is therefore not considered here. By considering that piezoelectric and deformation potential coupling are of the same order of magnitude in the low-density case as shown for $\mathrm{MoS_2}$ \cite{Kaasbjerg13}, we approximate $|D_{1,ac}^{(\mathrm{DP})}|^2\approx\frac{1}{2}|D_{1,ac}^{(\mathrm{total})}|^2$. These arguments are in line with Ref.~\cite{SeligDiss}. Additionally DFT results only obtain the modulus of the matrix element, such that the sign has to be chosen. As indicated by DFT calculations \cite{Rasmussen15} and measurements \cite{He13}, we choose the relative sign between conduction and valence band to be negative. This is important for exciton-phonon coupling since in Eq.~(\ref{eq_Xph_coupling}) the difference of electron and hole coupling enters.

\end{document}